\documentclass[a4paper,11pt]{article}
\pdfoutput=1 
\usepackage{jheppub} 

\usepackage[utf8]{inputenc}

\usepackage{comment}
\usepackage{graphicx,rotating,amsmath,amsfonts,amssymb}
\usepackage{color}

\newcommand{\ie}{{i.e.~}}
\newcommand{\eg}{{e.g.~}}

\newcommand{\be}{\begin{equation}}
\newcommand{\ee}{\end{equation}}
\newcommand{\br}{\begin{eqnarray}}
\newcommand{\bea}{\begin{eqnarray}}
\newcommand{\eea}{\end{eqnarray}}
\newcommand{\er}{\end{eqnarray}}
\newcommand{\ba}{\begin{array}}
\newcommand{\ea}{\end{array}}
\newcommand{\bi}{\begin{itemize}}
\newcommand{\ei}{\end{itemize}}
\newcommand{\bn}{\begin{enumerate}}
\newcommand{\en}{\end{enumerate}}
\newcommand{\bc}{\begin{center}}
\newcommand{\ec}{\end{center}}

\newcommand{\eV}{{\rm eV}}

\newcommand{\td}{\mathrm{d}}
\newcommand{\diag}{\mathrm{diag}}

\newcommand{\eps}{\epsilon}

\title{The Evolving Planck Mass in Classically Scale-Invariant Theories}

\author[a]{K. Kannike,}
\author[a]{M. Raidal,}
\author[a]{C. Spethmann}
\author[a]{and H. Veerm\"ae}

\affiliation[a]{National Institute of Chemical Physics and Biophysics,  R\"avala 10, 10143 Tallinn, Estonia}

\emailAdd{kristjan.kannike@cern.ch}
\emailAdd{martti.raidal@cern.ch}
\emailAdd{christian.spethmann@kbfi.ee}
\emailAdd{hardi.veermae@cern.ch}

\abstract{We consider classically scale-invariant theories with non-minimally coupled scalar fields, where the Planck mass and the hierarchy of physical scales are dynamically generated. The classical theories possess a fixed point, where scale invariance is spontaneously broken. In these theories, however, the Planck mass becomes unstable in the presence of explicit sources of scale invariance breaking, such as non-relativistic matter and cosmological constant terms. We quantify the constraints on such classical models from Big Bang Nucleosynthesis that lead to an upper bound on the non-minimal coupling and require trans-Planckian field values. We show that quantum corrections to the scalar potential can stabilise the fixed point close to the minimum of the Coleman-Weinberg potential. The time-averaged motion of the evolving fixed point is strongly suppressed, thus the limits on the evolving gravitational constant from Big Bang Nucleosynthesis and other measurements do not presently constrain this class of theories. Field oscillations around the fixed point, if not damped, contribute to the dark matter density of the Universe. }

\begin{document} 
\maketitle
\flushbottom

%%%%%%%%%%%%%%%%%%%%%%%%%%%%%%%%%%%%%%%%%%%%%%%%%%%%%%%%%%
\section{Introduction}
\label{sec:in}
%%%%%%%%%%%%%%%%%%%%%%%%%%%%%%%%%%%%%%%%%%%%%%%%%%%%%%%%%%

The discovery of a light Standard Model-like Higgs boson at the LHC~\cite{Chatrchyan:2012xdj,Aad:2012tfa}, and the apparent absence of any stabilisation mechanism that might protect its mass against radiative corrections from higher scales, have led to an uncomfortable situation in elementary particle theory. As a result, there has recently been a renewed interest in scalars in classically scale-invariant theories~\cite{Bardeen:1995kv,Hempfling:1996ht,Chang:2007ki,Foot:2007as,Foot:2007ay,Meissner:2006zh,Foot:2007iy,Iso:2009nw,Iso:2009ss,AlexanderNunneley:2010nw,Foot:2010av,Holthausen:2009uc,Foot:2010et,Ishiwata:2011aa,Lee:2012jn,Heikinheimo:2013fta,Heikinheimo:2014xza,Hur:2011sv,Carone:2013wla,Chivukula:2013xka,Chun:2013soa,Englert:2013gz,Farzinnia:2013pga,Hambye:2013sna,Holthausen:2013ota,Iso:2012jn,Khoze:2013oga,Khoze:2013uia,Allison:2014zya,Antipin:2013exa,Davoudiasl:2014pya,Dermisek:2013pta,Farzinnia:2014xia,Gabrielli:2013hma,Hashimoto:2013hta,Hill:2014mqa,Kannike:2014mia,Kannike:2015kda,Khoze:2014xha,Kubo:2014ida,Kubo:2014ova,Lindner:2014oea,Radovcic:2014rea,Allison:2014hna,Altmannshofer:2014vra,Ametani:2015jla,Benic:2014aga,Carone:2015jra,Das:2015nwk,Endo:2015ifa,Endo:2015nba,Farzinnia:2015uma,Foot:2011et,Guo:2014bha,Guo:2015lxa,Haba:2015rha,Humbert:2015epa,Kang:2014cia,Kang:2015aqa,Karam:2015jta,Oda:2015gna,Okada:2015gia,Pelaggi:2014wba,Plascencia:2015xwa,Sannino:2015wka,Wang:2015sxe,Ahriche:2015loa,Ahriche:2016cio,Ahriche:2016ixu,Das:2016zue,Farzinnia:2015fka,Ghorbani:2015xvz,Haba:2015lka,Haba:2015nwl,Kubo:2015cna,Haba:2015qbz,Helmboldt:2016mpi,Ishida:2016ogu,Jinno:2016knw,Kannike:2016bny,Karam:2016rsz,Khoze:2016zfi,Marzola:2016xgb,Wang:2015cda,Wu:2016jdo,Hatanaka:2016rek,Karananas:2016kyt}, which might offer a new approach to the question of the origin and co-existence of different scales in Nature.

In a classically scale-invariant theory, all mass scales such as the QCD scale or the electroweak scale must be generated dynamically. Furthermore, if the theory should also contain gravity, the Planck mass itself must be generated through the dynamical breaking of scale invariance~\cite{Minkowski:1977aj,Zee:1978wi,Smolin:1979uz,Adler:1980bx,Wetterich:1987fm,Wetterich:2002wm,Shaposhnikov:2008xb,Lin:2014mua,Salvio:2014soa,Kannike:2015apa,Einhorn:2015lzy,Einhorn:2016mws,Ferreira:2016vsc,Cooper:1982du,Finelli:2007wb,Tronconi:2010pq,Cerioni:2010ke,Kamenshchik:2012rs,Bezrukov:2012hx,GarciaBellido:2011de}. 
Such models with scale-invariant quartic potentials of several scalar fields, 
all  non-minimally coupled to gravity, possess non-trivial properties already at the classical level.
The simplest model capable of generating a hierarchy of scales comprises two scalar fields and has been recently studied in Refs.~\cite{Shaposhnikov:2008xb,Bezrukov:2012hx,GarciaBellido:2011de,Ferreira:2016vsc}. 
It contains an interesting feature, namely that the two-dimensional field space has a dynamically stable direction, which is related through a global Weyl transformation to the massless Goldstone mode of the spontaneously broken scale symmetry --- the dilaton. It is conceivable that the hierarchical field values in such a fixed point can be linked to the existence of hierarchical mass scales in Nature.

The stability of the fixed point depends critically on the scale invariance of the theory. In the observable Universe, scale invariance is broken by the abundance of massive particles and by any bare cosmological constant term. In case the breaking is explicit, there are two distinct effects. First, if the breaking happens in a sector which is only gravitationally coupled to the scalars, the fixed point, and with it the dynamically generated Planck mass, changes in time, giving rise to observable effects. As explicit breaking necessarily introduces a new scale, then the observable quantity independent of the choice of units is the ratio of the Planck scale to that scale. Second, if the breaking directly modifies the scalar potential, \eg~through quantum effects, a minimum of the potential is generated, which tends to stabilise the Planck mass.\footnote{It has been argued~\cite{Shaposhnikov:2008xi,Ghilencea:2015mza,Ghilencea:2016ckm} that it is also possible to construct scale-invariant renormalisation schemes. In this work, however, we consider dimensional transmutation induced by quantum effects, such as the dynamical QCD 
or Coleman-Weinberg~\cite{Coleman:1973jx} scales, to be the manifestation of breaking the scale invariance.} The first aim of this paper is to study how the existence of Standard Model matter, dark matter and dark energy affects the stability of the theory, to quantify the effects of the evolving Planck scale and to derive the corresponding constraints on the theory parameters. 

The fixed point of the model \cite{Ferreira:2016vsc} only determines the ratio of the field values  of the scalars, but does not fix the absolute scale which evolves as discussed previously. 
It is well known that the Coleman-Weinberg~\cite{Coleman:1973jx} mechanism  can generate a minimum on the flat direction via dimensional transmutation. 
Our second aim is to study whether and how the Coleman-Weinberg mechanism operates in such a setup, and quantify the phenomenological consequences. 
We show that a new stable point is, indeed, generated which is displaced from the minimum of the scalar potential due to the presence of space-time curvature.
In comparison to the classical case, the rate at which the stable point evolves in time is heavily suppressed, solving the Planck scale stability problem.
We note that this scenario requires the introduction of a bare cosmological constant term, and the related hierarchy problem cannot be addressed in the present context alone.
Interestingly,  scalar oscillations around the stable point may contribute to the dark matter density of the Universe.

The paper is organised as follows. In Section~\ref{sec:gen} we revisit the classically scale invariant scenario with several scalar fields non-minimally coupled to gravity and 
expand it by  adding extra  matter to the Lagrangian.   In Section~\ref{sec:one} we study the resulting cosmology and the changing Planck mass in the case of one non-minimally coupled scalar field. In Section~\ref{sec:two} we return to the problem of breaking scale invariance in the case of two non-minimally coupled scalars. In Section~\ref{sec:pheno} we study the phenomenological bounds on a changing Planck mass. We conclude in Section~\ref{sec:end}.

%%%%%%%%%%%%%%%%%%%%%%%%%%%%%%%%%%%%%%%%%%%%%%%%%%%%%%%%%%
\section{Planck Mass from the Spontaneous Breaking of Scale Invariance}
\label{sec:gen}
%%%%%%%%%%%%%%%%%%%%%%%%%%%%%%%%%%%%%%%%%%%%%%%%%%%%%%%%%%

In this paper we consider theories with the action 
\begin{align}\label{eq:S_gen}
	S 
	= \int \td^{4}x \sqrt{-g}\left(\frac{1}{2} M^{2}(\phi_i) R - \frac{1}{2}\sum_{i}(\nabla\phi_{i})^2 - V_{\phi}(\phi_i) + \mathcal{L}_{\rm M} \right),
\end{align}
where the effective Planck mass $M^{2}(\phi_i)$ is a function of the fields, $R$ denotes the Ricci scalar,  $V_{\phi}$ is the potential of the scalar fields $\phi_{i}$ and $g = \det{(g_{\mu\nu})}$ is the determinant of the metric. We use natural units $\hbar = c = 1$ and the metric signature $(-+++)$. The matter Lagrangian $\mathcal{L}_{\rm M}$ describes the matter content of the Universe in the broadest sense, \ie~including dark matter, radiation, and possibly a bare cosmological constant term, as well as non-derivative interactions with the scalar fields $\phi_i$. We generically do not require the matter Lagrangian $\mathcal{L}_{\rm M}$ to be scale-invariant. The variation of the action \eqref{eq:S_gen} results in the equations of motion
\begin{align}
	\label{eom_g0}
	M^{2} G_{\mu\nu}	
&	=  T^{\xi}_{\mu\nu} +  T^{\phi}_{\mu\nu} + T^{\rm M}_{\mu\nu},
	\\
	\label{eom_phi0}
	\square \phi_i	
&	= 	\frac{\partial V_{\phi}}{\partial \phi_i} 
	-	\frac{1}{2} \frac{\partial M^{2}}{\partial \phi_i}  R
        -      \frac{\partial \mathcal{L}_{\rm M}}{\partial \phi_i},
\end{align}
where the stress-energy tensors corresponding to the dynamical Planck mass and the scalar fields are given by
\begin{align}
	T^{\xi}_{\mu\nu} 
&	\equiv \left(\nabla_{\mu}\nabla_{\nu} - g_{\mu\nu}\square\right)M^{2},
	\\
	T^{\phi}_{\mu\nu} 
&	\equiv \sum_i \left(\nabla_{\mu} \phi_{i}\nabla_{\nu} \phi_{i} - \frac{1}{2} g_{\mu\nu}  (\nabla \phi_{i})^{2} \right)
	- g_{\mu\nu} V_{\phi},
\end{align}
and $T^{\rm M}_{\mu\nu}$ is the stress energy of the remaining degrees of freedom.

Using the scalar field equations \eqref{eom_phi0} together with the identity $\nabla^{\mu} \phi_{i}\nabla_{\mu} \phi_{i} = \frac{1}{2}\square \phi_i^{2} - \phi_i \square \phi_i$, it is possible to rearrange the trace of Einstein equations as
\begin{align}
	\frac{1}{2}\square\left(6 M^{2} + \phi^{2}\right)	
&	=	\frac{1}{2}\left(2M^{2} - \phi \frac{\partial M^{2}}{\partial \phi}\right) R
 	+	\left(\phi \frac{\partial V_{\phi}}{\partial \phi} - 4 V_{\phi}\right) 
 	+	\left(T^{\rm M} - \phi \, \frac{\partial \mathcal{L}_{\rm M}}{\partial \phi} \right),
	\label{eq:tr:Einstein:eqs}
\end{align}
where $\phi = \sqrt{ \sum_i \phi_{i}^{2}}$ is the radial direction in the field space. For further convenience, we define
\begin{equation} 
\delta V \equiv \phi \frac{\partial V_{\phi}}{\partial \phi} - 4 V_{\phi} \,,
\label{eq:deltaV}
\end{equation}
which quantifies any deviations from a scale-invariant potential. If the potential is classically scale-invariant, then a non-vanishing $\delta V$ arises from conformal anomalies.

The identity $\phi\, \partial X/\partial \phi = c X$ implies that the function $X$ is homogeneous of degree $c$ in the fields $\phi_i$. The right-hand side (RHS) of equation \eqref{eq:tr:Einstein:eqs} vanishes if the Planck mass is homogeneous of degree 2, the potential is homogeneous of degree 4, and the matter sector has only scale-invariant interactions. In other words, the RHS vanishes if the action is scale-invariant. In this case the combination
\begin{equation}\label{def:sigma}
	\sigma \equiv \frac12 \left(6 M^{2} + \phi^{2} \right)
\end{equation} 
obeys the equation of motion of a free, minimally coupled, massless scalar degree of freedom and thus its behaviour is known: any initial oscillating behaviour of $\sigma$ will be damped as $a^{-4}$ in a an expanding Universe. Thus the field eventually relaxes to a constant, thereby setting the vacuum in which scale invariance is spontaneously broken. In this fixed point the value of $\sigma$ can be any positive real number, thus implying an infinitely degenerate vacuum. The dynamics of the oscillations of $\sigma$ around this vacuum indicates that  $\sigma$ is related to the massless Goldstone boson corresponding to spontaneously broken scale invariance, as will be further explained in Section \ref{sec:two:si}. 

Note that $J_\mu \equiv \partial_\mu \sigma$ is the Noether current corresponding to scale symmetry~\cite{Oda:2013uca}. Vanishing of the the RHS of equation \eqref{eq:tr:Einstein:eqs} is thus implied by the Noether theorem, since in a classically scale invariant scenario $J_\mu$ is conserved, $\nabla^\mu J_\mu = 0$.  On the other hand, if scale invariance is violated explicitly, \ie if $\nabla^\mu J_\mu \neq 0$, then fixed point described above does generally not exist. As a result $\sigma$ and therefore also the Planck mass will evolve. 

At this point we would like to briefly comment on the issue of evolving dimensional fundamental constants~\cite{Duff:2002vp}. It can be argued that time variation of such a constant can as well be a result of time variation of our choice of units and thus it is not physical. In fact, if we decided to work in the Einstein frame, then the Planck mass would be constant by construction. To demonstrate that the evolution of the Planck scale is physically meaningful, we need a physical reference scale to fix the unit. In case of explicit breaking such a scale (\eg the QCD scale) always exists. Thus in the Einstein frame where the units are fixed, so that the Planck mass remains constant, we would observe the evolution of the physical reference scale. We also refer the reader to the discussion at the end of Sec.~\ref{sec:two:si}.

All explicit mass terms in the matter Lagrangian $\mathcal{L}_{\rm M}$ contribute to the trace $T^{\rm M}$. The only such term in the Standard Model is the Higgs mass parameter, and all fundamental fermion masses are propotional to the Higgs vacuum expectation value (VEV). If the Higgs mass parameter arises from the spontaneous breaking of scale invariance, \ie~from the non-minimally coupled scalar field obtaining a VEV, then none of the masses of the Standard Model  fundamental fermions contribute to the right-hand side of equation \eqref{eq:tr:Einstein:eqs}. However, baryon masses still contribute because they are dominantly generated by strong dynamics, and baryons make up most of the density of the visible matter of the present Universe. 
Finally, also dark matter may contribute to the right-hand side of equation \eqref{eq:tr:Einstein:eqs} if its dynamics is not scale-invariant as happens in models with strong dynamics in the dark sector, \eg dark technicolour~\cite{Heikinheimo:2013fta,Heikinheimo:2014xza,Hur:2011sv}.                         

In the following we restrict ourselves to a scale-invariant classical potential
\begin{align}\label{eq:V_phi}
	V_{\phi} = \frac{1}{4}\sum_{ij}\lambda_{ij}\phi_{i}^{2}\phi_{j}^{2},
\end{align}
and the effective Planck mass
\begin{align}\label{eq:M_phi}
	M^{2} \equiv \sum_{i}\xi_{i} \phi_{i}^2,
\end{align}
so that the Lagrangian has an additional $\mathbb{Z}_2$ symmetry for each field $\phi_{i}$. For simplicity, below we shall assume that no interaction terms between the scalars and the matter fields exist,
\begin{align} \label{eq:no_interactions}
       \frac{\partial \mathcal{L}_{\rm M}}{\partial \phi_i} = 0,
\end{align}
unless stated otherwise.

The dynamics of a single field is usually studied in the Einstein frame where gravitational and quantum physics are cleanly separated. We will, however, stay in the Jordan frame because it is not generally possible to canonically normalise more than one non-minimally coupled scalar field in the Einstein frame \cite{Kaiser:2010ps}. The two frames are related by a choice of units of length. In the Jordan frame we choose the (explicit) masses of particles as the measuring sticks for length and allow the field-dependent Planck mass to evolve. In the Einstein frame the units are tied to a constant Planck mass and the explicit mass terms acquire a field dependence. 

The choices \eqref{eq:V_phi} and \eqref{eq:M_phi} yield a scale-invariant model and thus a massless $\sigma$-field. This invariance is, however, broken by the quantum corrections due to the renormalisation group running of the couplings. We identify the renormalisation scale $\mu$ with the radial direction $\phi$ in the field space which is the standard procedure to RG-improve the effective potential. Plugging the tree level renormalisation group equation (RGE)-improved potentials into \eqref{eq:tr:Einstein:eqs} now give
\begin{align}\label{ddM}
	\square \, \sigma	
&	=	
	-	\frac{1}{2}\sum_{i}\beta_{\xi_{i}} \phi_{i}^2 \,R
 	+	\frac{1}{4}\sum_{ij}\beta_{\lambda_{ij}} \phi_{i}^2 \phi_{j}^2
 	+	T^{\rm M} ,
\end{align}
where $\beta_{\xi_{i}}$ and $\beta_{\lambda_{ij}}$ denote the $\beta$-functions of the corresponding coupling constants. The deviation from classical scale invariance $\delta V$ of the potential is given by the trace anomaly and related to the RGE running. The result for $\delta V$ is the same as from direct computation of the trace anomaly \cite{Callan:1970ze,Polchinski:1987dy,Coleman:1970je} (see e.g. \cite{Fortin:2011sz} for a recent overview). In the rest of the paper we will assume that the running of the non-minimal couplings contributes sub-dominantly to the breaking of scale invariance when the values of scalar couplings are small \cite{Kannike:2015apa}, and set $\beta_{\xi_{i}} = 0$. Note that at two-loop the Eq.~\eqref{ddM} can acquire additional contributions \cite{Jack:1990eb}.

An important consequence of the gravitational interaction, and especially the non-minimal coupling in this scenario, is that the stationary solution, if it exists, is not any more determined by the minimum of the potential, but by the vanishing of the scale symmetry breaking terms, \ie the RHS of \eqref{ddM}. In any cosmological epoch that is not dominated by the cosmological constant or radiation the Ricci scalar $R$ will generally evolve in time. The vanishing of the RHS of \eqref{ddM} will still determine a solution that might be followed approximately, given that the change of $R$ is slow enough.

 The $\sigma$-field \eqref{def:sigma} now reads explicitly
\begin{align}
	\sigma = \frac{1}{2}\sum_{i}(1+6\xi_{i}) \phi_{i}^2.
\end{align}
It follows that conformally coupled fields, for which $\xi_i = -1/6$, do not contribute.

To determine the fixed point, we solve the equations of motions for constant fields $\phi_{i} = v_{i}$. To obtain a compact equation, we eliminate $R$ from the scalar field equations \eqref{eom_phi0} and find
\begin{align}\label{Lambda0}
	\Lambda 
	= \frac{1}{4\xi_{i}v_{i}}\left.\frac{\partial V_{\phi}}{\partial \phi_{i}}\right|_{\phi_{i} = v_{i}}\,,
\end{align}
with the geometry being determined by $R = 4\Lambda$, where $\Lambda$ is a dynamically generated cosmological constant. We stress that solutions obeying this equation may exist only in an explicitly scale-invariant setting. In general, we have two qualitatively different scenarios for the fixed point:
\begin{enumerate}
	\item The potential is scale-invariant, \ie $\delta V_{\phi}$ as defined in Eq.~\eqref{eq:deltaV} is identically zero, and in a fixed point $T^{\rm M} = 0$. The cosmological constant is induced by the potential according to \eqref{Lambda0}, and the matter sector might contain a radiation component.
	\item The potential is not scale-invariant. Therefore $\delta V_{\phi} = 0$ is an additional constraint on the vacuum expectation values, and it is also required that $\dot{T}^{\rm M} = 0$. In this setup the matter sector may contain a bare cosmological constant term and possibly also a radiation component.
\end{enumerate}
In the following we study these scenarios in the special cases of one and two non-minimally coupled scalar fields.

%%%%%%%%%%%%%%%%%%%%%%%%%%%%%%%%%%%%%%%%%%%%%%%%%%%%%%%%%%
\section{One-Field Case}
\label{sec:one}
%%%%%%%%%%%%%%%%%%%%%%%%%%%%%%%%%%%%%%%%%%%%%%%%%%%%%%%%%%

Many qualitative features can already be observed in the single field scenario, where the Planck mass is simply $M^2 = \xi \phi^2$. Various cosmological aspects of this model, including the evolution of the Planck mass, have been studied in Refs.~\cite{Cooper:1982du,Finelli:2007wb,Tronconi:2010pq,Cerioni:2010ke,Kamenshchik:2012rs}. In a Friedmann-Robertson-Walker 
 Universe with a metric $g_{\mu\nu} =  \diag(-1, \delta_{ij} a^{2})$, the cosmological evolution is given by a modified Friedmann equation
\begin{align}\label{1field_f_eq}
	3\left(1 + \kappa\right) H^{2} M^{2} &	=  \rho,
\end{align}
where $\rho = \rho_{\phi} + \rho_{\rm M}$ is the total energy density and $\rho_{\phi} = \frac{1}{2} \dot{\phi}^{2} + V_{\phi}$ is the energy density of the scalar field. We have quantified deviations from the standard $\rm \Lambda CDM$ cosmology by a dimensionless parameter
\begin{align}\label{def:kappa}
	\kappa \equiv \frac{\partial_{t}M^{2}}{H M^{2}} = -H^{-1}\frac{\dot{G}}{G}\,,
\end{align}
 that gives the relative rate of change of the gravitational constant $G$ during one Hubble time. It measures how the gravitational constant scales with the scale factor $a$, \ie locally $G \propto a^{-\kappa}$. Due to phenomenological considerations we require that $\kappa \ll 1$ throughout cosmological history from Big Bang Nucleosynthesis (BBN) to the present.

As the second equation of motion we choose Eq.~\eqref{ddM} that can be recast as 
\begin{align}\label{1field_M2_eq}
	\left(\partial_{t}^{2} + 3H \partial_t\right)M^{2}	
&	= -\frac{2\xi}{1 + 6 \xi}\left(\delta V_{\phi} +  T^{\rm M}\right),	
\end{align}
where $\delta V$ is defined in Eq.~\eqref{eq:deltaV}. For simplicity we have neglected the usually sub-dominant running of the non-minimal coupling constant, \ie $\beta_{\xi} = 0$.

%%%%%%%%%%%%%%%%%%%%%%%%%%%%%%%%%%%%%%%%%%%%%%%%%%%%%%%%%%
\subsection{Scale-Invariant Scalar Sector}
\label{sec:one:si}
%%%%%%%%%%%%%%%%%%%%%%%%%%%%%%%%%%%%%%%%%%%%%%%%%%%%%%%%%%

Without quantum corrections $\delta V = 0$.  First, if $T_{\rm M} = 0$, then $\dot{M} = 0$ (or equivalently $\kappa = 0$) is a solution to the equations of motion. Moreover, a constant Planck mass is an attractor because Hubble friction drives all time derivatives to zero. As a result, the field $\phi$ obtains a VEV, and therefore this set-up spontaneously breaks scale invariance. In the case when the trace of the energy momentum tensor does not vanish, $T_{\rm M} \neq 0$, we instead observe a slowly evolving Planck mass.

In the following we derive the rate of change of the Planck mass \eqref{def:kappa} in terms of the energy density. This is the main result of this section. Note that the Planck mass follows a second order differential equation and its rate of change is therefore dependent on the initial conditions. Because of Hubble friction, however, the asymptotic result is insensitive to the initial rate of change of the Planck mass.  In order to simplify the derivation we represent time by the number of $e$-folds $N = \ln a$. In this case 
\begin{equation}
g_{\mu\nu} =  \diag(-H^{-2}(N), \delta_{ij} \exp (2N))\,,
\end{equation}
where the Hubble parameter $H(N)$ is the dynamical variable. As the gravitational sector is classically scale-invariant, $M^{2} = \xi \phi^{2}$ and $V = \lambda \phi^4/4 = \lambda \xi^{-2} M^{4} /4$. The equations of motion \eqref{1field_f_eq} and \eqref{1field_M2_eq} in terms of the Planck mass now read
\begin{align}
	3 (M^{2} + \partial_N M^{2}) H^{2}
	& =  \rho ,
	\\
	\partial_N^{2} M^{2} + \left(3 + \partial_N \ln H \right)\partial_N M^{2}	
	& = \frac{2\xi}{1 + 6 \xi}\frac{- T^{\rm M}}{H^{2}}
\end{align}
with the energy density of the scalar fields  given by 
\begin{equation}
\rho_{\phi} = \frac{\kappa^2}{8 \xi} H^2 M^2   + V_{\phi} .
\end{equation}
Notice that $-T_{\rm M} >0$ because of our choice of the metric signature.

The Hubble parameter $H$ appears algebraically and can be eliminated, thereby removing one equation. Introducing $\kappa$ into the equations by
\begin{align}
	\partial_N M^{2} = \kappa M^{2} ,
\end{align}
leaves us with a first order differential equation for $\kappa$: 
\begin{align}
\partial_N\kappa + \left[ \frac{3}{2}(1-\omega) + \frac{1}{2}\kappa - \frac{1}{2} \partial_N\ln(1 + \kappa)\right] \kappa
&	= \frac{6\xi}{1 + 6 \xi}\frac{- T^{\rm M}}{\rho}(1+ \kappa).	
\end{align}
where we defined an effective equation of state parameter $\omega$ by $\partial_{N}\rho \equiv -3(1+\omega) \rho$. The energy density $\rho=\rho_M+\rho_\phi$ also contains contributions from the scalar field. Therefore, the right-hand side is implicitly dependent on the dynamics of the Planck mass. We restrict out assumption to the set-up satisfying the following mild conditions:
\begin{enumerate}
	\item The gravitational constant is changing slowly, \ie $\kappa \ll 1$. Phenomenological constraints on the evolving Planck mass (see Sec.~\ref{sec:pheno}) imply that this condition is satisfied through the observable history of the Universe. 
	\item The energy density is independent of $\kappa$. In particular, this implies that the kinetic energy of the scalar field that determines the Planck mass does not give the dominant contribution to the energy density because $\kappa$ quantifies the time dependence of this field.\footnote{Formally, in the expansion $\rho = \rho_0 + \rho_1 \kappa + \mathcal{O}(\kappa^2)$ it is sufficient to require that the $\rho_0$ term is dominant. In fact, in most cases this is already guaranteed by the first condition, $\kappa \ll 1$.} 
\end{enumerate}
We then arrive at the linearised equation
\begin{align}
	\partial_N\kappa +\frac{3}{2}(1-\omega)\kappa
&	= \frac{6\xi}{1+6\xi} \frac{- T^{\rm M}}{\rho} + \mathcal{O}(\kappa^2),
\end{align}
where for consistency the source term on the RHS is of $\mathcal{O}(\kappa)$. The dependence on the initial value of $\kappa$ is lost after a few $e$-folds, the exact time depending on $\omega$.
After this initial transient period, $\kappa$ evolves according to the  asymptotic solution of the linearised equation,
\begin{align} \label{eq:kappa_int}
	\kappa(N) = \frac{6\xi}{1+6\xi} \int^{N}_{-\infty} \td N' \exp\left(- \frac{3}{2}(1-\omega) (N - N')\right) \frac{- T^{\rm M}}{\rho} .
\end{align}
The dominant components of the energy density and the trace of the stress-energy tensor might arise from different sources. The integral can be expressed in closed form if $\rho = \rho_{0} \exp(-3(1+\omega) \, N)$ and $T_{\rm M} =  T_{0} \exp(-3(1+\omega_{T})\, N)$, \ie if the total energy density has a constant equation of state parameter $\omega$. A similar effective equation of state parameter $\omega_{T}$ can also be attributed to the trace of the stress-energy tensor. The corresponding solution is given by 
\begin{align}
	\kappa = \frac{6\xi}{1+6\xi} \, \frac{1}{\frac{3}{2}(1+\omega) - 3\omega_{T} } \frac{- T^{\rm M}}{\rho}  .
        \label{eq:kappa}
\end{align}
We note the following cases:
\begin{enumerate}
	\item Assume, that the non-scale invariant sector dominates, \ie $\rho \approx \rho_{\rm M}$ and $\omega \neq 1/3$. In that case the trace of the stress-energy tensor is determined by the dominant matter component, $T_{M} = -\rho (1-3 \omega)$. This implies
\begin{align}\label{eq:kappa_1}
	\kappa = \frac{4\xi}{1+6\xi} \, \frac{1-3 \omega}{1-\omega}.
\end{align}
	A similar scenario was studied in~\cite{Cooper:1982du} as a perturbation of around the de Sitter background. 
	
	\item If $\rho \approx \rho_{\rm M}$ but $\omega = 1/3$, \ie the Universe is in a radiation dominated epoch, then the trace of the stress-energy tensor $T_{M}$ is determined by a subdominant component and there is an additional suppression due to the ratio $T^{\rm M}/\rho$, implying
\begin{align}
	\kappa = \frac{6\xi}{1+6\xi} \, \frac{1}{2 - 3\omega_{T} } \frac{- T^{\rm M}}{\rho} .
        \label{eq:kappa_BBN}
\end{align}

	\item If we drop the assumption that $\rho \approx \rho_{\rm M}$, then it is possible that the dominant contribution has $\omega \neq 1/3$ but originates from the scale-invariant sector. In this case the stress-energy tensor $T^{\rm M}$ will result from a subdominant component, and Eq.~\eqref{eq:kappa} with the suppression factor $T^{\rm M}/\rho$ can not be simplified further. An example of this scenario is provided by the two-field model discussed in the next section. The equations of motion have a stationary solution with a constant Planck mass that predicts that we will observe a vacuum energy dominated Universe with $\omega \approx -1$, where the vacuum energy is determined by the fields, \ie $\rho \approx \rho_{\phi}$.

\end{enumerate}
A numerical study of the cosmological solutions of the scale invariant model comprising two fields and  a cold matter component ($\omega = 0$) will be presented in 
Section~\ref{sec:two}.  The numerical analysis demonstrates that the analytical results obtained in this section can also be adapted to a scenario with many fields, provided that only one field gives the dominant contribution to the Planck mass.

%%%%%%%%%%%%%%%%%%%%%%%%%%%%%%%%%%%%%%%%%%%%%%%%%%%%%%%%%%
\subsection{Breaking Scale Invariance via the Coleman-Weinberg Mechanism}
\label{sec:one:bsi}
%%%%%%%%%%%%%%%%%%%%%%%%%%%%%%%%%%%%%%%%%%%%%%%%%%%%%%%%%%

If the scale invariance is broken explicitly by $\delta V \neq 0$, the situation is qualitatively different. As we will show shortly, the $\sigma$-field  obtains a mass. The system exhibits a time-dependent fixed point, corresponding to the Planck mass $M_{*}^{2} = \xi \phi_{*}^{2}$. We find that the rate of change of $M_{*}$ is now supressed relative to the scale-invariant case.

The ``fixed point'' is evolving in time. Therefore the field value will not follow it exactly, but instead oscillate around it. If those oscillations are not depleted by the decay or annihilation of the scalar, they will scale as non-relativistic matter and therefore contribute to the dark matter abundance. 

In this section we will focus on the case where scale invariance is explicitly broken by quantum corrections. Without specifying the running of the scalar self-coupling, at one-loop level the RGE improved potential is given by
\begin{align}\label{eq:V_1field}
	V_{\phi} (\phi)
&	= \frac{1}{4}\lambda(\phi) \phi^{4} + V_{0},
\end{align}
where $V_{0}$ is a bare vacuum energy. Without loss of generality all contributions to the vacuum energy are assumed to be absorbed into $V_{0}$. 
The field value $\phi_{*}$ at the fixed point is obtained by setting the RHS of Eq.~\eqref{1field_M2_eq} to zero:
\begin{align}
	\frac{1}{4}\beta(\phi_{*}) \phi_{*}^{4} - 4V_{0} + T_{\rm M} = 0.
\label{eq:vac:cond}
\end{align}
Because $T_{\rm M}$ evolves in time, $\phi_{*}$ is not stationary. The temporal average of the field $\phi$ will still trace $\phi_{*}$ quite accurately if $T_{\rm M}$ changes slowly compared to the other scales that determine the dynamics of $\phi$, such as its mass. Explicitly, from Eq.~\eqref{eq:vac:cond} we obtain the rate of change $\kappa$ corresponding to $M_*^2$,
\begin{align}
	\kappa= \frac{\partial_N M_*^2}{M_*^2} = \frac{2 \xi^2}{\beta(\phi_{*})} \, \frac{-\partial_N T_{\rm M}}{ M_*^4} ,
\label{eq:kappa_bsi}
\end{align}
where, as before, $N=\ln a$.
We see that the rate of change of $M_*$ is now heavily suppressed by the ratio of the energy density of the Universe over the Planck density, $\rho/M_*^4$, compared to the scale-invariant case in Eq.~\eqref{eq:kappa}. 

The vacuum energy density $\rho_{\Lambda}$  is defined by the value of the potential at the vacuum expectation value of the field, \ie by the field value in the absence of other matter, $v \equiv \phi_{*}(T_{\rm M} = 0)$. Then \eqref{eq:vac:cond} implies
\begin{align}\label{rho_lambda}
	\rho_{\Lambda} 
	\equiv V_{\phi}(v) 
	= \left( 1+4\frac{\lambda(v)}{\beta(v)}\right) V_{0} .
\end{align}
We choose a model-independent approach and do not specify the origin of the $\beta$-function.

The observed value of the cosmological constant today is of the order $\rho_{\Lambda} = \mathcal{O}(10^{-3} \eV)^{4}$, while $V_{0}$ is presumably close to the Planck density. Since $V_{0} \gg \rho_{\Lambda}$ the hierarchy between $\rho_{\Lambda}$ and $V_{0}$ introduces fine tuning. Especially we see that a fine-tuned cosmological constant requires $\lambda(v) < 0$ implying that the potential has a non-trivial minimum at $v_{0}$, around which the coupling constant can be expanded as
\begin{align}
	\lambda(\phi) = \beta(v_{0})\left( -\frac{1}{4} + \ln \frac{\phi}{v_{0}} \right).
	\label{eq:lambda:lin:approx}
\end{align}
Comparing this with the expression for $\lambda(v)$ from Eq.~\eqref{rho_lambda} yields
\begin{align}
	\frac{v}{v_{0}} = 1 + \frac{1}{4}\frac{\rho_{\Lambda}}{V_{0}} + \mathcal{O}\left(\frac{\rho_{\Lambda}}{V_{0}}\right)^{2}.
\end{align}
We see that the cosmological constant is equivalent to a VEV $v$ that is slightly displaced from the minimum of the potential $v_0$. We stress that a non-vanishing VEV is possible even if the value of the potential at its minimum is zero, as long as $\beta(v) > 0$. 

\begin{figure}[tb]
\begin{center}
\includegraphics[width=0.45\textwidth]{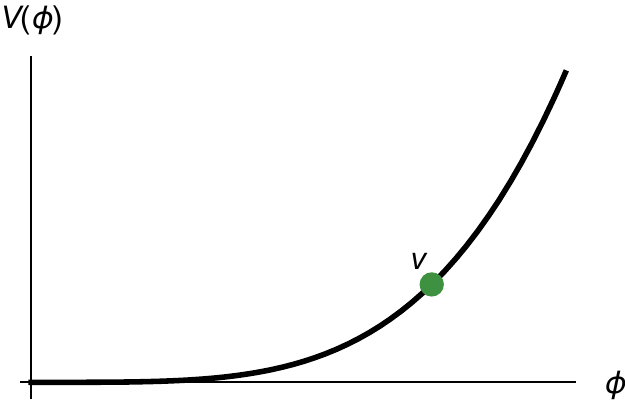}
\includegraphics[width=0.45\textwidth]{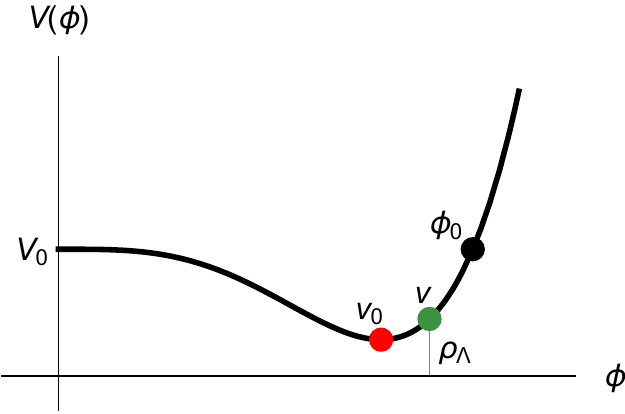}
\caption{Schematic depiction of the tree-level potential (left panel) with the fixed point in green. The potential with quantum corrections is shown at right. The fixed point $v$ is displaced from the Coleman-Weinberg minimum $v_{0}$ due to spacetime curvature. In the linear approximation \eqref{eq:lambda:lin:approx}, the coupling $\lambda_{\phi}$ runs through zero at $\phi_{0}=e^{1/4} \, v_0$.}
\label{fig:schematic:V}
\end{center}
\end{figure}

A schematic depiction of the potential is displayed in Fig.~\ref{fig:schematic:V}. The left panel shows the tree-level potential with the fixed point $v$. In the right panel, the true fixed point $v$ (green) is slightly displaced from the Coleman-Weinberg minimum of the potential $v_{0}$ (red).

For completeness, we proceed to estimate the size of the fluctuations around the minimum. To this aim we work in the approximation $V_{0} \gg \rho_{\Lambda}$ and set $v=v_{0}$. The deviation of the effective Planck mass from the equilibrium $\delta_{M}  \equiv M^{2} - M_{*}^{2}$ evolves according to \eqref{1field_M2_eq},
\begin{align}
	\left(\partial_{t}^{2} + 3H \partial_t\right)\delta_{M}
	\approx \omega_{M}^{2} \delta_{M},
\end{align}
where we neglected the time derivatives of $M_{*}$ and defined the oscillation frequency
\begin{align}
	 \omega_{M} = M_{*}\,\sqrt{\frac{-\beta(\phi_{*})}{\xi(1 + 6 \xi)}} .
\end{align}
The energy density corresponding to these oscillations scales as $a^{-3}$; they therefore behave as non-relativistic matter, and in principle this component  may contribute to the dark matter abundance. The value of $\rho_{\phi}$ depends on the initial conditions and also on possible interactions with other degrees of freedom.  In practice, such interactions can be expected to exist and the abundance of the $\phi_{i}$ fields might be depleted by decay or annihilation processes. As there is no empirical data on the initial conditions or interactions, we treat $\rho_{\phi}$ as an additional input parameter. 

Using the $\kappa$ parameter defined in Eq.~\eqref{def:kappa}, we recast the energy density of the scalar field as
\begin{align}
	\bar{\rho}_{\phi} 
	= \frac{H^{2}M^{2}}{8\xi} \kappa^{2} + \bar{V}_{\phi} .
\end{align}
Above we defined
$
	\bar{V}_{\phi} (\phi)
	\equiv V_{\phi} (\phi) - \rho_{\Lambda},
$
so that in the vacuum $\bar{V}_{\phi} = 0$. In this way the cosmological constant is separated from the energy density of the field $\rho_{\phi}$. The parameter $\kappa$ oscillates with the opposite phase relative to the Planck mass. It obtains its largest value when the energy is given purely by the kinetic term. Therefore, with the help of Eq.~\eqref{1field_f_eq}, we conclude that the envelope of oscillations is 
\begin{align}
	\kappa_{\rm osc}^{2} = \frac{8 \xi}{3}\frac{\bar{\rho}_{\phi}}{\rho},
\end{align}
where $\rho$ is the total energy density. It follows that the canonical Friedmann equations with $\kappa_{\rm osc} \ll 1$ correspond to $|\xi| \ll 1$, or to a negligible contribution of the scalar field to the total energy density, $\bar{\rho}_\phi \ll \rho$.

%%%%%%%%%%%%%%%%%%%%%%%%%%%%%%%%%%%%%%%%%%%%%%%%%%%%%%%%%%
\section{Two-Field Case}
\label{sec:two}
%%%%%%%%%%%%%%%%%%%%%%%%%%%%%%%%%%%%%%%%%%%%%%%%%%%%%%%%%%

%%%%%%%%%%%%%%%%%%%%%%%%%%%%%%%%%%%%%%%%%%%%%%%%%%%%%%%%%%
\subsection{Scale-Invariant Scalar Sector}
\label{sec:two:si}
%%%%%%%%%%%%%%%%%%%%%%%%%%%%%%%%%%%%%%%%%%%%%%%%%%%%%%%%%%

Consider now two real fields $\phi_1$ and $\phi_2$ obeying the action 
\begin{align}\label{eq:S_phi}
	S 
	= \int \td^{4}x \sqrt{-g}\left(\frac{1}{2} \left( \xi_{1} \phi_{1}^2 + \xi_{2} \phi_{2}^2 \right) R - \frac{1}{2}\sum_{i}(\nabla\phi_{i})^2 - V_{\phi} \right) + S_{\rm M}, 
	\end{align}
with a quartic potential of the form \eqref{eq:V_phi},
\begin{align}
	V_{\phi} = \frac{1}{4}\left( \lambda_{11} \phi_{1}^{4} + 2\lambda_{12} \phi_{1}^{2} \phi_{2}^{2} + \lambda_{22} \phi_{2}^{4} \right),
\end{align}
and a large hierarchy between the self couplings, $\lambda_{11} \ll |\lambda_{12}| \ll \lambda_{22}$. It was recently observed that, neglecting RGE running and the matter content of the Universe, the model has a stable fixed direction determined by the ratio \cite{Ferreira:2016vsc}
\begin{align}\label{vev_ratio1}
	\frac{\phi_{2}^{2}}{\phi_{1}^{2}} = \frac{\lambda_{12}\xi_1 - \lambda_{11}\xi_2}{\lambda_{12}\xi_2 - \lambda_{22}\xi_1}.
\end{align}
This ratio is easily obtained from Eq.~\eqref{Lambda0}. Note that the condition is scale-invariant as it only fixes the ratio. At the stationary point, the model predicts a cosmological constant
\begin{align}\label{vev_Lambda}
	\Lambda \equiv \frac{V_{\phi}}{M^{2}} = \frac{1}{4} \frac{ \lambda_{11}\lambda_{22} - \lambda_{12}^{2} }{\lambda_{11}\xi_2 - \lambda_{12}\xi_1} \: \phi_{2}^{2},
\end{align}
and Planck mass
\begin{align}\label{vev_M}
	M^{2} 
&	= 4\Lambda \; \frac{\lambda_{22}\xi_1^{2} - 2\lambda_{12}\xi_1\xi_2 + \lambda_{11}\xi_2^{2}}{\lambda_{11}\lambda_{22} - \lambda_{12}^{2}}.
\end{align}
A large hierarchy between the Planck mass and the cosmological constant therefore implies $\lambda_{12}^{2} \approx \lambda_{11}\lambda_{22}$, rendering the potential \eqref{eq:V_phi} almost a perfect square. 

\begin{figure}[tb]
\begin{center}
\includegraphics[width=0.475\textwidth]{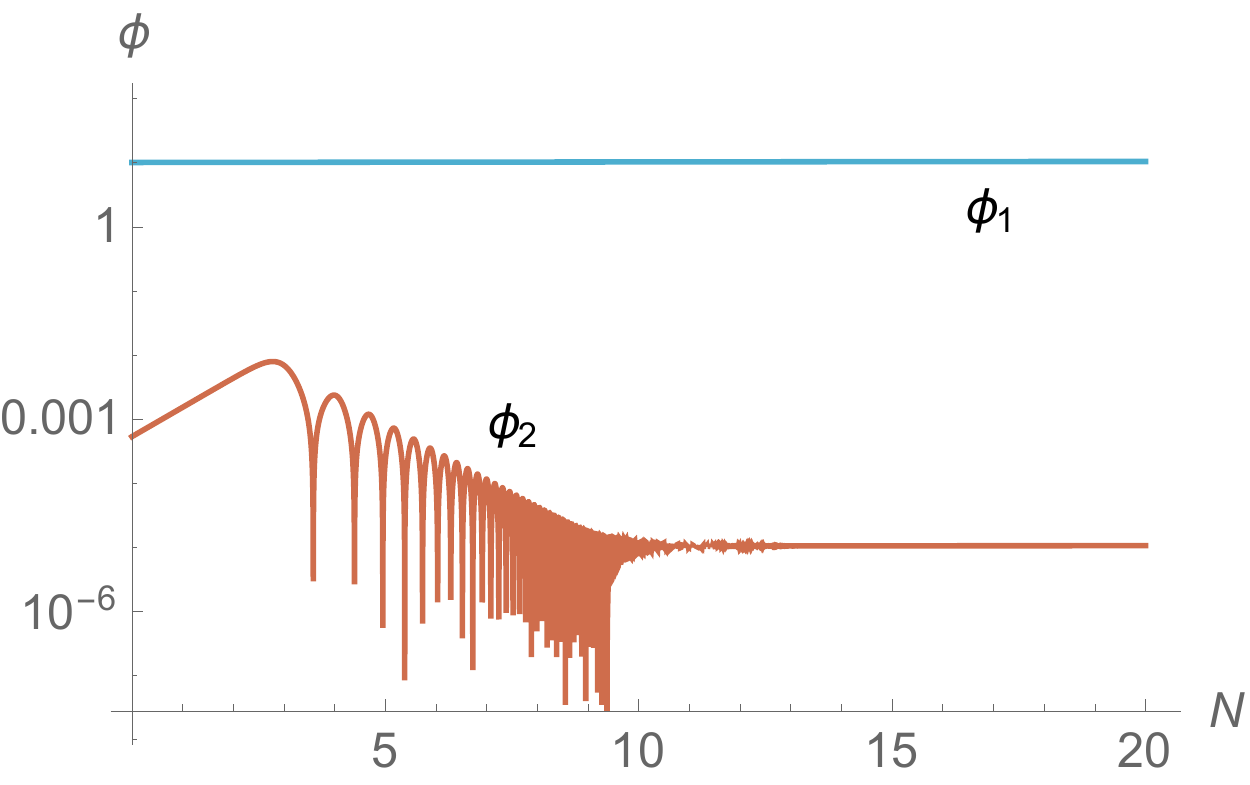}
\hfill
\includegraphics[width=0.475\textwidth]{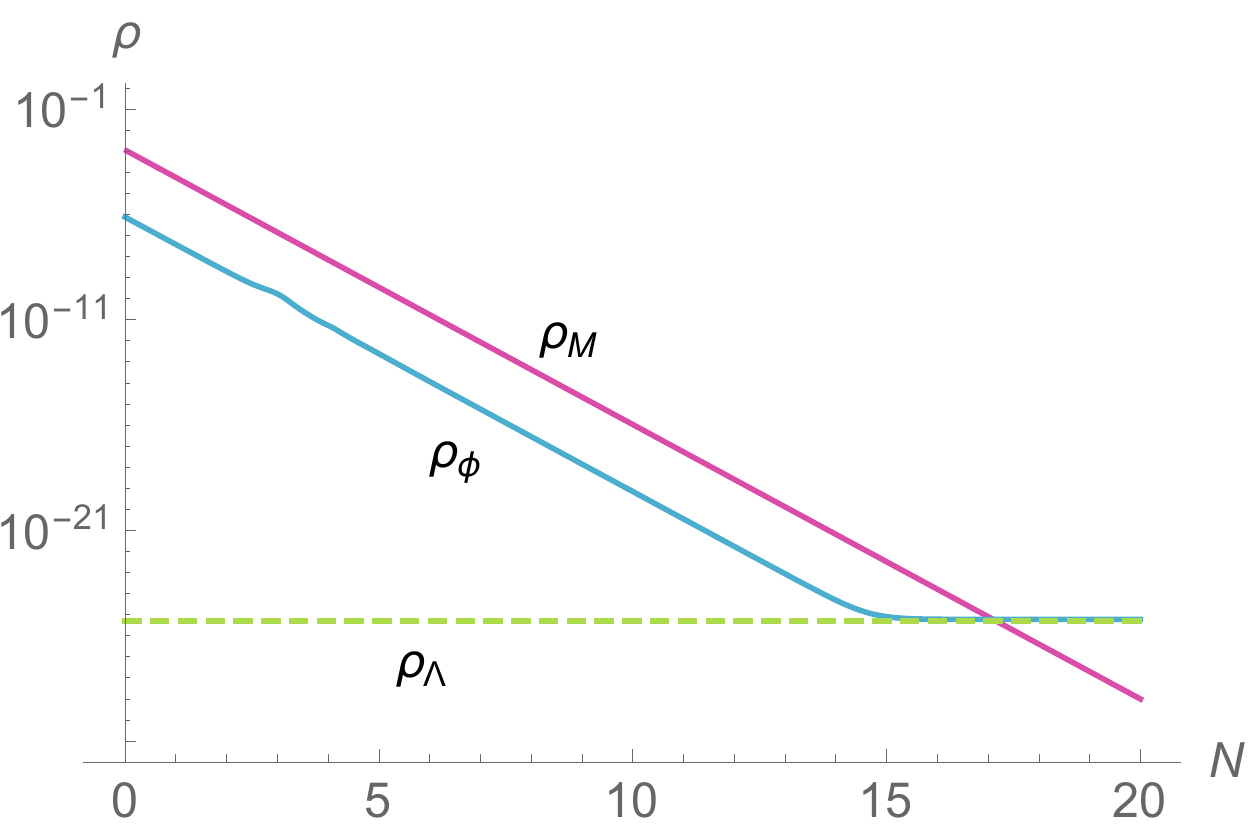}\\[2ex]

\includegraphics[width=0.475\textwidth]{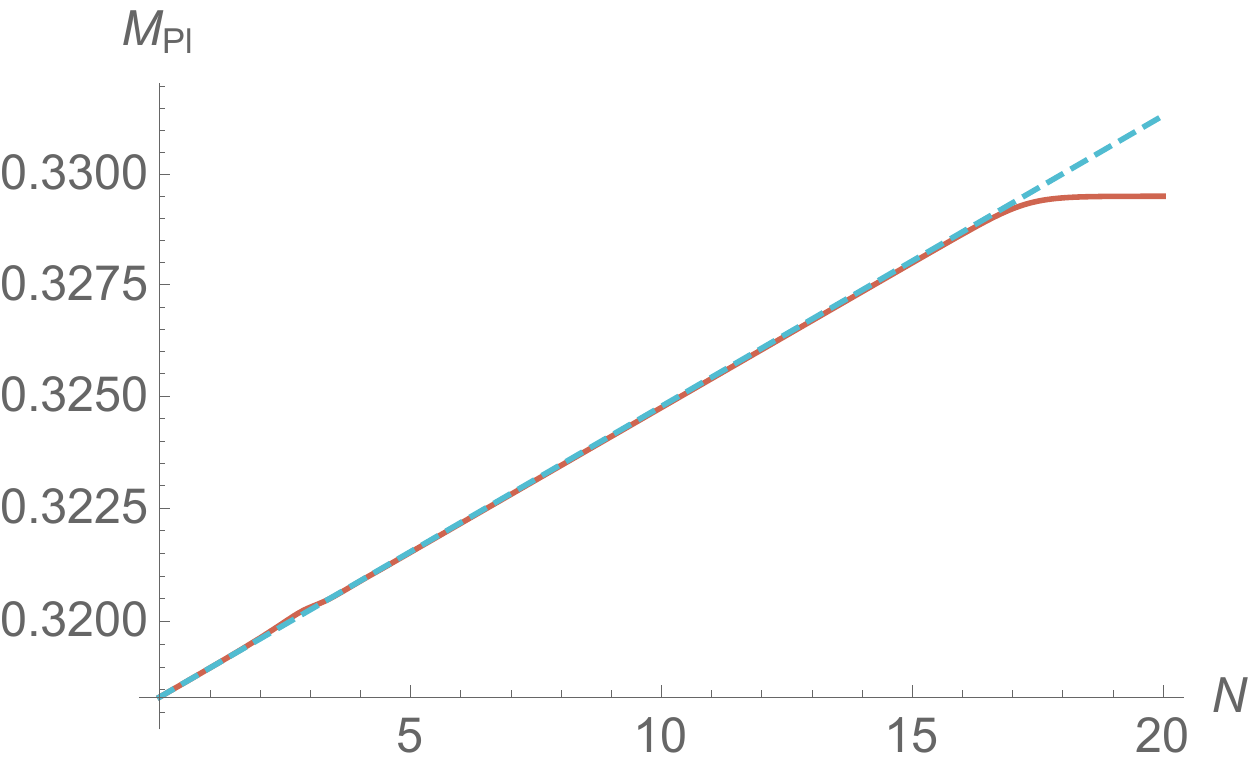}
\hfill
\includegraphics[width=0.475\textwidth]{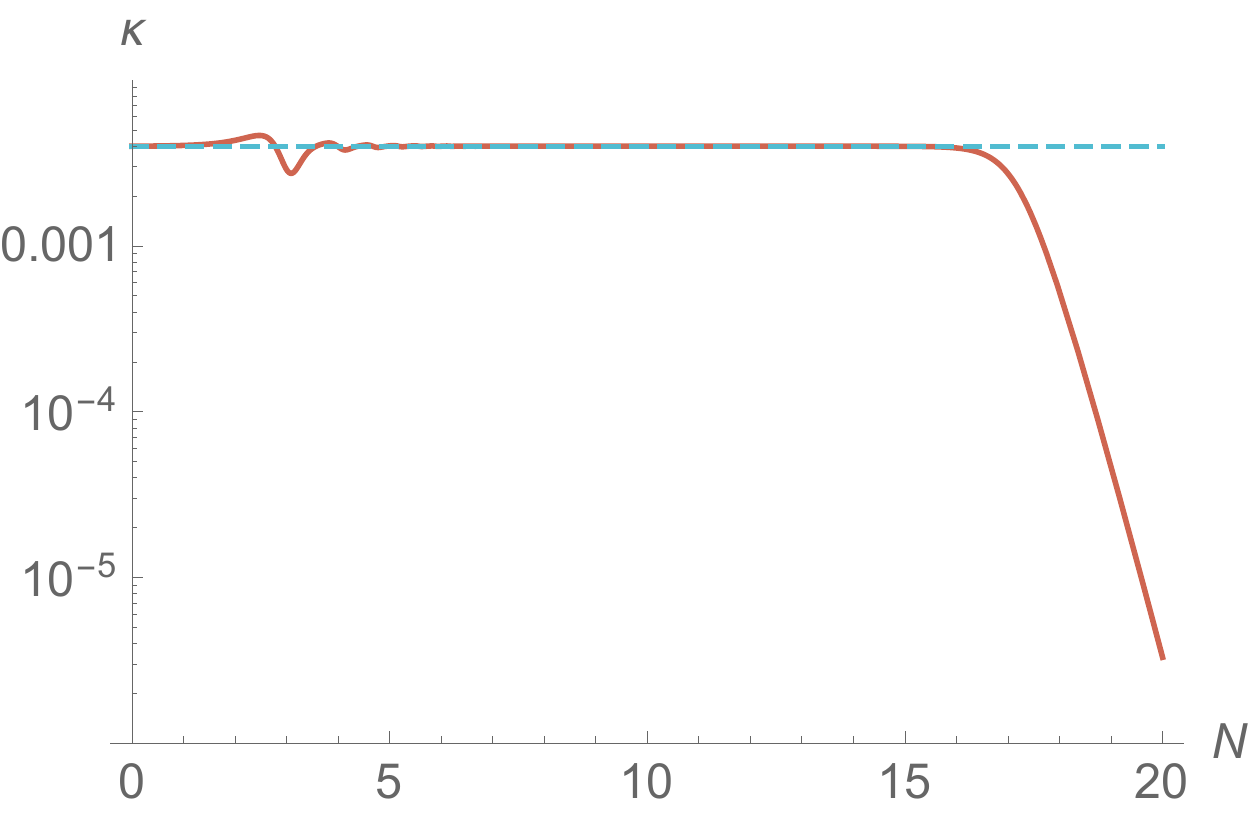}

\caption{A numerical study of the cosmological solutions of the scale invariant model \eqref{eq:S_phi} with a cold matter component ($\omega = 0$).
Top-left: The evolution of fields $\phi_{1}$ (blue) and $\phi_{2}$ (red).
Top-right: The evolution of the energy densities of the scalar field $\rho_\phi$ (red), cold matter $\rho_M$ (blue) and cosmological constant $\rho_{\Lambda}$ from the scalar fields \eqref{vev_Lambda} (green).
Bottom-left: The numerical evolution of the Planck mass (solid line) compared to the analytic approximation $M^{2} \propto \exp \kappa$ evaluated for the matter dominated epoch (dashed line).
Bottom-right: The numerical evolution of the $\kappa$ parameter (solid line) compared to its analytic approximation \eqref{eq:kappa_1} evaluated for the matter dominated epoch (dashed line). Parameters used for the plot: $\lambda_{11} = 10^{-25}$, $\lambda_{22} = 0.1$, $\lambda_{12} = -0.9999 \times 10^{-13}$, $\xi_1 = 10^{-3}$, $\xi_2 = 0.9$. The horizontal axis denotes time in $e$-folds, $N = \ln a$.
}
\label{fig:nonCW}
\end{center}
\end{figure}

Due to the the hierarchy of couplings, the Planck mass at the fixed point is mainly sensitive to changes in the field $\phi_1$. It is therefore expected that the rate of change of the Planck mass derived for the one field case \eqref{eq:kappa} can be adapted to the current scenario by simply identifying $\xi$ with $\xi_1$. This gives
\begin{align}
	\kappa \approx \frac{6\xi_1}{1+6\xi_1} \, \frac{1}{\frac{3}{2}(1+\omega) - 3\omega_{T} } \frac{- T^{\rm M}}{\rho}  .
\end{align}
In Fig.~\ref{fig:nonCW} we show results of a numerical study of cosmological solutions for the scale invariant model \eqref{eq:S_phi} with an additional cold matter component ($\omega = 0$) that breaks scale invariance explicitly. In all the panels one can observe the transition from the matter dominated epoch to the cosmological constant dominated epoch at $N \approx 17$, as is explained in the caption of the figure. In particular, the top-left panel demonstrates how the fields settle into to the fixed point, while the
top-right panel demonstrates the onset of vacuum energy dominance.
 It is evident from figure that the energy density of the fields is diluted in the same way as the energy density of cold matter, since the slopes of the corresponding curves are identical. This behaviour indicates that $\phi_{2}$ obtained a mass from the spontaneous symmetry breaking of scale invariance, i.e. from the nearly constant value of $\phi_{1}$.

In the bottom-left and bottom-right panels of Fig.~\ref{fig:nonCW}  we present the corresponding numerical results for the behaviour of 
the Planck mass and the $\kappa$ parameter. Those results are compared against the predictions of 
the one-field scenario \eqref{eq:kappa_1} in the matter dominated epoch, presented by the dashed lines.
One can observe that the one-field approximation works well. The cosmological constant \eqref{vev_Lambda} in this scenario arises purely from spontaneous breaking of scale invariance and therefore does not contribute to $\kappa$. This is indicated by the rapid decline of the $\kappa$ parameter at the onset of the cosmological constant dominated epoch.

A special case with a vanishing tree-level cosmological constant, \ie when $\lambda_{12} = -\sqrt{\lambda_{22} \lambda_{11}}$, which has been studied in~\cite{Shaposhnikov:2008xb}. In that case the fixed direction \eqref{vev_ratio1} simplifies to
\begin{equation}\label{eq:v1:v2}
		\frac{\phi_{2}^{2}}{\phi_{1}^{2}}
	=	\sqrt{\frac{\lambda_{11}}{\lambda_{22}}} \equiv \eps\,,
\end{equation}
and the potential \eqref{eq:V_phi} can be recast as 
\begin{align}\label{eq:Vphi_eps}
	V_{\phi} = \frac14 \lambda_{22}\left(\phi_{2}^{2} - \eps\, \phi_{1}^{2}\right)^{2}.
\end{align}
We find that the ratio \eqref{eq:v1:v2} coincides with the minimum of the potential (flat direction) and is independent of the non-minimal couplings. The latter is easily understandable by noting that in the absence of a cosmological constant the vacuum spacetime geometry is Minkowski. 
It is most easily seen from Eq.~\eqref{eq:v1:v2} that the hierarchy between the dimensionless coupling constants $\lambda_{11}$ and $\lambda_{22}$ translates into the hierarchy between the scales $v_2$ and $v_1$, \eg $\lambda_{11} \ll \lambda_{22}$ implies $v_1 \gg v_2$. The induced hierarchy persists in the presence of a non-vanishing cosmological constant \eqref{vev_Lambda}.  In the rest of this section we require that the hierarchy in the non-minimal couplings be smaller than the hierarchy of the quartic couplings, in particular $\eps^{2} \equiv \lambda_{11}/\lambda_{22}  \ll  \xi_1^{2}/\xi_2^{2}$, such that the larger of the field values gives the dominant contribution to the Planck mass. 

From the potential \eqref{eq:Vphi_eps} it follows that as $\phi_{1}$ gets a VEV, $\phi_{2}$ obtains the mass
\begin{equation}\label{eq:mass_phi2}
	m_{\phi_2}^{2} = \sqrt{\lambda_{11} \lambda_{22}} \phi_{1}^{2}\,.
\end{equation}
This behaviour can in fact be witnessed in Fig.~\ref{fig:nonCW}. However, in case $\Lambda \neq 0$, the mass will also get contributions from the cosmological constant and the above formula will only hold approximately~\cite{GarciaBellido:2011de}.

To illuminate the role of the massless degree of freedom, let us consider the Einstein frame. To this aim we simplify the two-field model further and neglect the contribution of $\phi_{2}$ to the Planck mass. So we set  $\lambda_{12} = -\sqrt{\lambda_{22} \lambda_{11}}$ and  $\xi_{2} = 0$  and obtain the action
\begin{align}\label{eq:S_phi_simpJ}
	S 
	= \int \td^{4}x \sqrt{-g}\left(\frac{1}{2} \xi_{1} \phi_{1}^2 R - \frac{1}{2}(\nabla\phi_{1})^2 - \frac{1}{2}(\nabla\phi_{2})^2  - \frac14 \lambda_{22} \left(\phi_{2}^{2}  - \eps \phi_{1}^{2}   \right)^2   \right).
\end{align}
The transition to the Einstein frame is carried out with the Weyl transformation 
\begin{align}
	\phi_{i}  &\to \bar{\phi}_{i} = \exp(-\omega/f) \phi_{i} \, ,
	\\
	g_{\mu\nu} &\to \bar{g}_{\mu\nu} = \exp(2\omega/f)  g_{\mu\nu} \, ,
\end{align}
where the local scaling factor is fixed by $\exp(\omega/f) = M(\phi_{1})/M_{\rm Pl}$. Here $M_{\rm Pl}$ and $f$ are dimensional constants. 
The scaling factor $\omega$ now effectively replaces the degree of freedom of the field $\phi_{1}$, and can be identified with the dilaton. Altogether, these expressions define a change of variables in which the action reads
\begin{align}\label{eq:S_phi_simpE}
	S 
	= \int \td^{4}x \sqrt{-\bar{g}}  \bigg[
&		\frac{1}{2} M_{\rm Pl}^2 \bar{R} 
		- \frac{1}{2} (\bar{\nabla}\omega)^2 - \frac{1}{2f^2} (\bar{\nabla}\omega)^2 \bar{\phi}_{2}^2
	\nonumber\\
&		- \frac{1}{2 f} \bar{\nabla}\omega \bar{\nabla}\bar{\phi}_{2}^{2}
		- \frac{1}{2} (\bar{\nabla}\bar{\phi}_{2})^2
		- \frac14 \lambda_{22} \left( \bar{\phi}_{2}^{2}  - \frac{\eps}{\xi_{1}} M_{\rm Pl}^2    \right)^2  
	\bigg],
\end{align}
where we made the identification $f = M_{\rm Pl} \sqrt{(1 + 6\xi_{1})/\xi_{1}}$. As expected of a Goldstone boson, $\omega$ is a massless and derivatively coupled degree of freedom, and scale invariance is replaced by the global shift symmetry $\omega \to \omega + \eta$. This symmetry generates the following Noether current
\begin{align}\label{eq:E_J}
	\bar{J}_{\mu} 
	= \left(1+ \frac{\bar{\phi}_{2}^2}{f^2}\right) \bar{\nabla}_{\mu} \omega + \frac{1}{2 f} \bar{\nabla}_{\mu} \bar{\phi}_{2}^2
	\propto \frac{1}{M(\phi_{1})^2}\bar{\nabla}_{\mu} \sigma \,.
\end{align}
The continuity equation $\bar{\nabla}^{\mu}\bar{J} _{\mu} = 0$ in the Einstein frame is equivalent to its counterpart in the Jordan frame, $\square \sigma = 0$ (see  Eq.~\eqref{ddM}). We note, however, that $\sigma$ does not satisfy a massless free wave equation in the Einstein frame. The current \eqref{eq:E_J} is a special case of the scale current found in~\cite{GarciaBellido:2011de}.

Based on the calculations in the Jordan frame, we argued that scale symmetry breaking would cause the Planck mass to evolve in time. 
Yet this would be impossible in the Einstein frame as the Planck mass $M^{2} = \xi_{1} v_{1}^{2}$ is constant by construction, thus what is the corresponding effect in the Einstein frame? 
To reconcile the two frames we note that if the Jordan frame contains an explicit mass $m$, then in the Einstein frame it is transformed into $m v_{1}/\phi_{1}$. A (slowly) evolving $\phi_{1}$ is observable in form of the evolution of the explicit mass $m$. The two frames are related by a local redefinition of the unit of length. A frame independent
formulation of the observable effects can therefore be obtained by comparing masses that explicitly break the scale invariance to the masses generated by the spontaneous breaking of scale invariance.

%%%%%%%%%%%%%%%%%%%%%%%%%%%%%%%%%%%%%%%%%%%%%%%%%%%%%%%%%%
\subsection{Breaking Scale Invariance via the Coleman-Weinberg Mechanism}
\label{sec:two:bsi}
%%%%%%%%%%%%%%%%%%%%%%%%%%%%%%%%%%%%%%%%%%%%%%%%%%%%%%%%%%

Up to now we have neglected the running of coupling constants. In the classical Gildener-Weinberg approach in flat spacetime \cite{Gildener:1976ih} one parametrises the fields as $\phi_{1} = \phi \cos \theta$, $\phi_{2} = \phi \sin \theta$, so that the scalar potential \eqref{eq:V_phi} takes the form 
\begin{equation}
  V_{\phi} = \frac{1}{4} \lambda(\phi, \theta) \phi^{4} \equiv \frac{1}{4} \left[ \lambda_{11}(\phi) \cos^{4} \theta + 2 \lambda_{12}(\phi) \cos^{2} \theta \sin^{2} \theta + \lambda_{22}(\phi) \sin^{4} \theta \right] \, \phi^{4} + V_0,
  \label{eq:V:polar}
\end{equation}
where we have identified the renormalisation scale $\mu$ with the scale $\phi$ and included a bare vacuum energy $V_0$ term as we did for the single field potential \eqref{eq:V_1field}. The angle $\theta = \theta_{0}$ follows  from the ratio \eqref{vev_ratio1}. On the other hand, the flat direction of the potential obtained from the equations $\lambda(\theta) = 0$, $\partial_{\theta} \lambda(\theta) = 0$ reads
\begin{equation}\label{eq:CW_flat}
  \lambda_{12} + \sqrt{\lambda_{11} \lambda_{22}} = 0, \quad \tan^{2} \theta_{0} = \sqrt{\frac{\lambda_{11}}{\lambda_{22}}}\,,
\end{equation}
where the first equation (with $\lambda_{12} < 0$) gives the necessary and sufficient condition for the potential to have a flat direction. We stress that the solution for $\tan^{2} \theta$ exactly matches Eq.~\eqref{eq:v1:v2} in the absence of the cosmological constant \eqref{vev_Lambda}. As noted before, this is not a coincidence since \eqref{eq:CW_flat} implies that spacetime is flat in the fixed point. Thus the non-minimal coupling does not contribute and the fields relax into the minimum of the potential. In all, the conditions \eqref{eq:CW_flat} and $\Lambda \approx 0$ imply that it is possible to use the flat spacetime prescription for the Coleman-Weinberg mechanism.

At one-loop level the beta functions $\mu \; \mathrm{d}\lambda_{ij}/\mathrm{d}\mu = \beta_{\lambda_{ij}}$ for the quartic couplings are given  by
\begin{align}
 (4 \pi)^{2} \beta_{\lambda_{11}} &= 2 (9 \lambda_{11}^{2} + \lambda_{12}^{2}),
  \label{eq:beta:11}
 \\
  (4 \pi)^{2} \beta_{\lambda_{12}} &= 2 \lambda_{12} (3 \lambda_{11} + 4 \lambda_{12} + 3 \lambda_{22}),
\label{eq:beta:12}
 \\
 (4 \pi)^{2} \beta_{\lambda_{22}} &= 2 (9 \lambda_{22}^{2} + \lambda_{12}^{2}).
 \label{eq:beta:22}
\end{align}
The potential is bounded from below if and only if the conditions $\lambda_{11} > 0$, $\lambda_{22} > 0$ and $\lambda_{12} + \sqrt{\lambda_{11} \lambda_{22}} > 0$ are satisfied. In essence, the Coleman-Weinberg mechanism relies on the fact that if one of these conditions is violated at low energies, a non-trivial minimum will be generated for the effective potential due to RGE running. We work in the region where $\lambda_{12} < 0$. In case of a large hierarchy $\lambda_{22} \gg |\lambda_{12}| \gg \lambda_{11}$, the $\beta$-functions indicate that the larger quartics run faster and thus, as we move towards the infrared, we will inevitably reach the point where the third stability condition, $\lambda_{12} + \sqrt{\lambda_{11} \lambda_{22}} > 0$, is violated. An energy scale where \eqref{eq:CW_flat} is satisfied therefore always exists.

Following the Gildener-Weinberg approach we now restrict our attention to the potential along the flat direction $\theta = \theta_{0}$ and write as before
\begin{align}
	V_{\phi} = \frac{1}{4}\lambda(\phi)\phi^{4} + V_0.
\end{align} 
In this way we have effectively reduced the problem to the one-field case discussed in Section~\ref{sec:one:bsi}. Nevertheless, the two-field scenario provides a minimal self-contained setting that provides the beta functions necessary for generating a CW minimum for the potential.

As long as the logarithm $\ln(\phi/\phi_{0})$ remains small, the effective coupling $\lambda$ can be approximated by\footnote{For completeness we provide an analytic approximation for the solution of the RGEs obtained by assuming a large hierarchy $\lambda_{22} \gg |\lambda_{12}| \gg \lambda_{11}$:
\begin{align*}
	\lambda_{22}(\mu) 
&	= \frac{\lambda_{22}(\mu_{*})}{1 - \frac{9}{8\pi2} \lambda_{22}(\mu_{*}) \ln\left(\frac{\mu}{\mu_{*}}\right)},
	\qquad
	\lambda_{12}(\mu) 
	= \lambda_{12}(\mu_{*}) \left(\frac{\lambda_{22}(\mu)}{\lambda_{22}(\mu_{*})}\right)^{1/3},
	\\
	\lambda_{11}(\mu) 
	&	= \lambda_{11}(\mu_{*}) 
	+ \frac{\lambda_{12}^{2}(\mu_{*})}{3\lambda_{22}(\mu_{*})}\left[ 1 - \left(\frac{\lambda_{22}(\mu)}{\lambda_{22}(\mu_{*})}\right)^{-1/3} \right],
\end{align*}
where $\mu_{*}$ is a reference scale. We checked that this approximation is in a very good agreement with the numerical solution of the exact RGEs and it also matches well with \eqref{lambda_eff} for $\ln \frac{\phi}{\phi_{*}} < \mathcal{O}(10)$.}
\begin{equation}\label{lambda_eff}
  \lambda(\phi) = \beta_{\lambda}(\phi_{0}) \ln \frac{\phi}{\phi_{0}},
\end{equation}
so that it runs through zero at the scale $\phi_{0}$. At one-loop level,
\begin{equation}
  \beta_{\lambda}(\phi_{0}) = \frac{\lambda_{11}(\phi_{0}) \lambda_{22}(\phi_{0})}{2 \pi^{2}}  = \frac{\lambda_{12}^{2}(\phi_{0})}{2 \pi^{2}}.
\end{equation}
As an approximation we will consider only the VEV of $\phi$ in a flat background, $v_{0} = e^{-\frac{1}{4}} \phi_{0}$. Because $\phi$ is the pseudo-Goldstone boson of classical scale invariance, its mass,
\begin{equation}
	m_{\phi}^{2} = \beta_{\lambda}(\phi_{0}) v_{0}^{2} = \frac{\lambda_{11} \lambda_{22}}{2\pi^2} v_{0}^{2},
\end{equation}
 is loop-suppressed.

The mass of the other mass eigenstate $\Phi$ is well approximated by the tree level result \eqref{eq:mass_phi2} that we repeat here for convenience,
\begin{equation}
	m_{\Phi}^{2} = \sqrt{\lambda_{11} \lambda_{22}} v_{0}^{2}.
\end{equation}
Because the mixing between the fields is negligible, the mass eigenstates are approximately $\phi \approx \phi_{1}$ and $\Phi \approx \phi_{2}$. We see that for fixed $\lambda_{22}$, increasing the hierarchy by decreasing $\lambda_{11}$ decreases the mass of $\phi$ and makes the Coleman-Weinberg minimum shallower.

As in the one-field case we find that the value of the potential in the minimum generates a vacuum energy contribution 
\begin{align}
	V(v_{0}) = V_{0} - \frac{1}{16} \beta_{\lambda} v_{0}^{4}\,,
\end{align}
with the wrong sign compared to the present experimentally determined vacuum energy density. Therefore, based on phenomenological considerations alone, the introduction of a bare cosmological constant term is necessary to cancel the negative contribution from the Coleman-Weinberg minimum.  

We conclude that the cosmological implications of the two-field model are the same as for the one-field model described in Section \ref{sec:one:bsi}. A non-zero cosmological constant corresponds to a fixed point which is displaced from the minimum of the potential. A large hierarchy between the bare vacuum energy and the observed cosmological constant can only be achieved by fine-tuning the couplings of the scalar fields.

%%%%%%%%%%%%%%%%%%%%%%%%%%
\section{Phenomenology}
\label{sec:pheno}
%%%%%%%%%%%%%%%%%%%%%%%%%%%

In this section we will compare the current experimental limits on a changing Planck mass with the predictions from our model, thereby establishing bounds on the non-minimal coupling of the scalar fields. We first consider the classical, scale-invariant scenario and then analyse how our conclusions are affected when quantum effects are taken into account.

The strongest limits on a possible time variation of the Planck mass can be derived from BBN, since any modification of the expansion history of the Universe changes the predicted primordial isotope abundances. We therefore calculate the expected rate of change of the Planck mass at the BBN epoch, where for concreteness we assume $T_{\rm BBN}=1$ MeV.

%%%%%%%%%%%%%%%%%%%%%%%%%
\subsection{The Scale-Invariant Scenario}
%%%%%%%%%%%%%%%%%%%%%%%%%

At the time of BBN the Universe is radiation dominated. The relevant case is therefore described by Eq.~\eqref{eq:kappa_BBN}, which we here repeat for convenience:
\begin{align}
	\kappa = - \, \frac{1}{H} \, \frac{\dot{G}}{G} = \frac{6\xi}{1+6\xi} \, \frac{1}{2 - 3\omega_{T} } \frac{- T^{\rm M}}{\rho} .
        \label{eq:kappa_BBN_repeat}
\end{align}
The upper limit on the left-hand side of Eq.~\eqref{eq:kappa_BBN_repeat} can be derived from primordial isotope abundances. The resulting upper bound on the relative change of the Planck mass is \cite{19953}
\begin{align} 
	\left| \frac{\dot{G}}{G} \right|_{\mathrm{BBN}} < 1.7 \cdot 10^{-13} \; \mbox{yr$^{-1}$} = 3.6 \cdot 10^{-45} \; \mbox{GeV} . 
\end{align}
The Hubble rate at $T=1$ MeV is given by 
\begin{align} 
H = 1.66 \, g_*^{1/2} \, \frac{T^2}{M_{\rm Pl}} = 4.5 \cdot 10^{-25} \; \mbox{GeV}, 
\end{align}
where $g_*=10.75$ is the number of relativistic degrees of freedom including the electron/positron plasma. The upper bound on $\kappa$ from BBN is therefore
\begin{align} 
| \kappa_{\rm BBN} | < 8.1 \cdot 10^{-21} . 
\label{eq:kappa_limit}
\end{align}

To evaluate the right-hand side of Eq.~\eqref{eq:kappa_BBN_repeat} we need to calculate the energy density and the trace of the stress-energy tensor during the BBN era, as well as the scaling parameter $\omega_T$ of the trace.\footnote{Here $\omega_T$ only serves to describe the scaling behaviour of the trace resulting from a sub-dominant component of the Universe, and is not identical to the fraction of the pressure over the energy density of this component.}
The temperature of 1 MeV corresponds to a redshift of $z=4 \cdot 10^9$. At this time the Universe is radiation-dominated with a total energy density of
\begin{align}
	\rho_{\mathrm{BBN}} = 9.4 \cdot 10^{-13} \; \mbox{GeV$^4$} . 
\end{align}
For a particle $X$ in thermal equilibrium, the contribution to the trace of the stress-energy tensor is 
\begin{align}
	T_{X}^{\rm M} = \frac{g_{X}m_{X}^2T^2}{2 \pi^2} \; xK_1(x),
	\label{eq:T_eq}
\end{align}
where $g_{X}$ is the number of degrees of freedom, $m_{X}$ is the mass of the particle, $x=m_{X}/T$, and $K_1$ is a Bessel function of the first kind. In the relativistic limit $x \to 0$ this expression simplifies to
\begin{align} 
	T_{X}^{\rm M} \approx \frac{g_{X}m_{X}^2T^2}{2 \pi^2} . 
\end{align}
The scaling parameter for such a particle in thermal equilibrium is
\begin{align}
	\omega_T = \frac{x \, K_2(x)}{3 \, K_1(x)} - 1, 
	\label{eq:omegaT_eq}
\end{align}
where $K_2$ is a Bessel function of the second kind. In the relativistic limit $\omega_T = - 1/3$.

For a non-relativistic relic (\ie out of equilibrium) component, the above equations do not apply. Instead, the trace of the stress-energy tensor is in this case identical to the energy density of the species, which is given by
\begin{align}
	T_{X}^{\rm M} = \rho_X = (z_{\rm BBN}-1)^3 \, \rho_{X,0}.
	\label{eq:T_noneq}
\end{align}
Here $z_{\rm BBN}$ is the redshift during BBN and $\rho_{X,0}$ is the energy density of the species today. For such a relic $\omega_T=0$.

The constraint on the non-minimal coupling $\xi$ of the model then depends on the particle species which makes the dominant contribution to the RHS of Eq.~\eqref{eq:tr:Einstein:eqs}. In the following we examine three different cases.

\begin{enumerate}

\item All Standard Model as well as dark matter particles contribute. This means that at $T=1$ MeV, the $e^+/e^-$ plasma dominates the trace of the stress-energy tensor. It is in thermal equilibrium with the photon bath, such that Eqns.~\eqref{eq:T_eq} and \eqref{eq:omegaT_eq} apply. In this case we find
\begin{align}
	T^{\rm M} & \approx T^{\rm M}_{e^+e^-} = 1.3 \cdot 10^{-13} \; \mathrm{GeV}^4, 
	\qquad \omega_T = - 0.24 .
\end{align}

\item If we assume that electrons acquire their mass through scale-invariant couplings to the scalar fields, their contribution to the right-hand side of Eq.~\eqref{eq:tr:Einstein:eqs} cancels. In this case the relevant contribution originates from the sum of the non-relativistic, out-of-equilibrium baryon and dark matter densities. Here Eq.~\eqref{eq:T_noneq} applies, such that
\begin{align}
	T^{\rm M} & \approx T^{\rm M}_{\mathrm{baryon}} + T^{\rm M}_{\mathrm{DM}} = 7.6 \cdot 10^{-18} \; \mathrm{GeV}^4, \qquad \omega_T = 0 .
\end{align}

\item Finally, if dark matter also acquires its mass through scale-invariant couplings to the scalar fields, only the baryon density contributes. In this case,
\begin{align}
	T^{\rm M} & \approx T^{\rm M}_{\mathrm{baryon}} = 1.2 \cdot 10^{-18} \; \mathrm{GeV}^4, \qquad \omega_T = 0 .
\end{align}

\end{enumerate}

The results for the constraints on the non-minimal coupling $\xi$ are presented in Table~\ref{tab:xi_limits}. In all three cases $\xi \ll 1$, requiring the VEV of the scalar field to be trans-Planckian. In the second line of the table we give the required VEV in units of the Planck mass. Even in the most optimistic case, the scalar field needs to acquire a VEV of order $10^7 \, M_{\rm Pl}$. 

\begin{table}
\begin{center}
\begin{tabular}{cclll}
\multicolumn{2}{c}{\textbf{Dominant contribution to $T^{\rm M}$}} & $e^+e^-$ plasma & baryons + DM & baryons \\ \hline
\boldmath${\delta V=0}$ & $\xi < $ \rule{0ex}{3ex} & $2.7 \cdot 10^{-20}$ & $3.3 \cdot 10^{-16}$ & $2.1 \cdot 10^{-15}$\\ 
 & $\langle \phi \rangle / M_{\rm Pl} > $ \rule{0ex}{3ex} & $6.1 \cdot 10^9$ & $5.5 \cdot 10^7$ &  $2.2 \cdot 10^7$ \\ \hline
\boldmath$\delta V \neq 0$ & $\xi/\sqrt{\beta} < $ \rule{0ex}{3ex} & $1.7 \cdot 10^{34}$ & $1.9 \cdot 10^{36}$ & $4.8 \cdot 10^{36}$
\end{tabular}
\end{center}
\caption{Experimental constraints on the evolution of Planck mass from Big Bang Nucleosynthesis. For the one-field scale-invariant case ($\delta V=0$), those set upper bounds on the non-minimal coupling $\xi$ and lower bounds on the scalar field $\langle \phi \rangle$, depending on which matter component makes the dominant contribution to the RHS of Eq.~\eqref{eq:tr:Einstein:eqs}.
If the scale invariance is broken by quantum effects ($\delta V \neq 0$), we derive upper bounds on the ratio $\xi/\sqrt{\beta}$, where $\beta$ is the $\beta$-function of the self-coupling. 
}
\label{tab:xi_limits}
\end{table}

%%%%%%%%%%%%%%%%%%%%%%%%%
\subsection{The Coleman-Weinberg Scenario}
%%%%%%%%%%%%%%%%%%%%%%%%%

The breaking of scale invariance through the Coleman-Weinberg mechanism tends to stabilise the Planck mass. The time-averaged parameter $\kappa$ is now given by Eq.~\eqref{eq:kappa_bsi},
\begin{align}
	\kappa= \frac{\partial_N M_*^2}{M_*^2} = \frac{2 \xi^2}{\beta(\phi_{*})} \, \frac{\partial_N T_{\rm M}}{ M_*^4}  = \frac{2 \xi^2}{\beta(\phi_{*})} \, \frac{3 (1+\omega_T) T_{\rm M}}{M_{\rm Pl}^4},\label{eq:kappa_bsi2}
\end{align}
where we used the scaling $\partial_N T_{\rm M} = 3 (1+\omega_T) T_{\rm M}$. Since in this case the value of the allowed couplings depends on the unknown $\beta$-function, we give the limits on the combination $\xi/\sqrt{\beta}$ in Table~\ref{tab:xi_limits}. Because of the huge suppression by the Planck density, all perturbative values for $\xi$ are allowed even for very small values of the $\beta$-function, and no trans-Planckian VEV is required for the scalar field.

%%%%%%%%%%%%%%%%%%%%%%%%%%%%%%%%%%%%%%%%%%%%%%%%%%%%%%%%%%
\section{Conclusions}
\label{sec:end}
%%%%%%%%%%%%%%%%%%%%

Classical scale invariance offers intriguing hints towards a solution to the hierarchy problem. In the physical Universe, however, scale invariance is broken. This breaking can be spontaneous or explicit. In this paper we investigated the effects of explicitly broken scale invariance on the stability of the Planck scale.  To this end, we identified the massless degree of freedom corresponding to spontaneously broken scale invariance.  We then investigated the effects of explicit breaking by an abundance of massive particles, a cosmological constant and quantum corrections in a toy model containing a single non-minimally coupled scalar field.  The results were extended to the classically scale-invariant scenario with two  hierarchical scalars~\cite{Shaposhnikov:2008xb,Ferreira:2016vsc}.

We found that, if the matter content receives masses from any other mechanism than the spontaneous breaking of the classical scale invariance, then
the fixed point in the space of the scalar fields --- which determines the Planck mass as well as the hierarchy of field values --- will evolve with time. 
The most stringent phenomenological constraints on an evolving Planck mass come from BBN. To find those constraints, we did not restrict ourselves to a particular 
phenomenological scenario but considered all logical possibilities for the different matter components of the Universe.
The resulting experimental bounds for different cases are listed in Table~\ref{tab:xi_limits}.
They constrain the non-minimally coupled scalar field value from below, and its non-minimal coupling to gravity from above.
The constraints are very stringent in all the cases requiring the field value to be several orders of magnitude above the Planck scale and the non-minimal
coupling to be very small. Our results imply that any possible identification of the studied system with physical fields in Nature is challenging both
phenomenologically and theoretically.

The cure lies in radiative corrections that generate a minimum in the flat direction via the Coleman-Weinberg mechanism. It is interesting that in the absence of a tree-level cosmological constant, the flat directions of the Coleman-Weinberg mechanism in Minkowski space-time and the fixed point of the mechanism of \cite{Ferreira:2016vsc} coincide. 
Thus dimensional transmutation via the Coleman-Weinberg mechanism is fully consistent and applicable in the scalar system under consideration.
As a result, a potential with a stable point is generated and the radial scalar field acquires a mass --- the scale invariance is broken explicitly
by radiative corrections. The stable point also evolves in time but this is strongly suppressed by the matter density over the Planck density.
Our resulting bounds on the non-minimal coupling $\xi$ in this case are also listed in the last row of Table~\ref{tab:xi_limits}. These bounds are so weak that they do not constrain the model in any meaningful way.

The Coleman-Weinberg scenario differs from the classical one in several additional aspects. In the classical two-field system the cosmological constant can be 
made much smaller than the Planck scale by adjusting dimensionless couplings only.  This corresponds to a potential that is almost a perfect square. On the other hand, the quantum 
case must necessarily involve a large bare vacuum energy, which must be fine-tuned against model parameters to obtain a small value for the 
observable cosmological constant. Second, it is phenomenologically interesting that in the quantum case the non-minimally coupled scalar field field may still oscillate around the stable point.
Cosmologically, these oscillations are diluted as non-relativistic matter, thus this scenario implies a new contribution to the dark matter abundance of the Universe.
We leave phenomenological studies of this possibility for future work.

%%%%%%%%%%%%%%%%%%%%%%%%%%%%%%%%%%%%%%%%%%%%%%%%%%%%%%%%%%

\section*{Acknowledgments}
We thank G. Ross and  D. Ghilencea for useful discussions and A. P\~{o}ldaru for careful reading.
This work was supported by the Estonian Research Council grants PUT716 and PUT799, 
the grant IUT23-6 of the Estonian Ministry of Education and Research, and by the EU through the ERDF CoE program TK133.

%===============================================================================
\bibliographystyle{JHEP}

\bibliography{infscales_refs}

\end{document}